\documentclass{physeauth}
\usepackage{graphicx}
\usepackage{amsmath}
\usepackage{amssymb}

\begin{document}

\begin{frontmatter}

\title{Phase measurements in open and closed Aharonov-Bohm interferometers}

\author[tau]{Amnon Aharony \thanksref{thank1}},
\author[tau]{Ora Entin-Wohlman} and
\author[wis]{Yoseph Imry}

\address[tau]{School of Physics and Astronomy, Sackler Faculty
of Exact Sciences, Tel Aviv University, tel Aviv 69978,
\\Department of Physics, Ben-Gurion University, Beer Sheva 84105,
Israel, \\and Argonne National laboratory, Argonne, IL60439, USA}
\address[wis]{Department of Condensed Matter Physics, Weizmann Institute
of Science, Rehovot 76100, Israel}

\thanks[thank1]{
Corresponding author. E-mail: aharony@post.tau.ac.il}

\begin{abstract}
Mesoscopic Aharonov-Bohm interferometers have been used in
attempts to measure the transmission phase of a quantum dot which
is placed on one arm of the interferometer. Here we review
theoretical results for the conductance through such
interferometers, for both the closed (two-terminal) and open
(multi-terminal) cases. In addition to earlier results for the
Coulomb blockade regime, we present new results for the strongly
correlated Kondo regime, and test the consistency of the two-slit
analysis of some data from open interferometer experiments.

\end{abstract}

\end{frontmatter}

\section{Introduction and Review of Experiments}

Mesoscopic quantum dots (QDs) represent artificial atoms with
experimentally controllable properties \cite{review,book,kastner}.
Connecting a QD via two one-dimensional (1D) `metallic' leads to
electron reservoirs, one can vary the attraction of electrons to
the QD by the `plunger gate voltage', $V_G$. Measurements of the
conductance ${\textbf G}$ through the QD, as function of $V_G$,
show peaks whenever the Fermi energy $\epsilon_F$ of the electrons
crosses a resonance on the QD.
 The quantum information
on the resonant tunnelling through the QD is contained in the {\it
complex} transmission amplitude, $t_{QD}=-i\sqrt{{\textbf
T}_{QD}}e^{i\alpha_{QD}}$.  It is thus of great interest to
measure the $V_G$-dependence of both the magnitude ${\textbf
T}_{QD}$ and the phase $\alpha_{QD}$.

The ``textbook" method to measure the phase of a wave is based on
the {\it two-slit interferometer} \cite{feynman}. In this
geometry, a coherent electron beam is split between two paths,
going through two slits, and one measures the distribution of
electrons absorbed on a screen behind the two slits. This
distribution contains information on the relative phases of the
electron wave functions in the two paths, via interference. In the
Aharonov-Bohm interferometr (ABI) \cite{azbel}, one adds a
magnetic flux $\Phi$ through the area between the two paths. This
adds the Aharonov-Bohm phase difference $\phi=e\Phi/\hbar c$
between the phases of the wave functions in the two branches of
the ring \cite{AB}.  Denoting the ``bare" transmission amplitudes
through each path by $t_i=|t_i|e^{i\alpha_i}$, the ``standard"
two-slit formula for the outgoing electron distribution, or
equivalently for the transmission through the ABI, has the form
\begin{eqnarray}
{\textbf
T}=|t_1e^{i\phi}+t_2|^2=A+B\cos(\phi+\beta),\label{twoslit}
\end{eqnarray}
with $\beta=\alpha_1-\alpha_2$ and
\begin{eqnarray}
A=|t_1|^2+|t_2|^2,\ \ \ \ B=2|t_1||t_2|.\label{selfcon}
\end{eqnarray}
However, as discussed below, this formula applies
only under very specific conditions.

Placing a QD on one path, and changing its plunger gate voltage
$V_G$, would vary the corresponding amplitude $t_1 \equiv t_{QD}$.
If the 2-slit formula were valid, it would allow the determination
of the dependence of $\alpha_{QD}$ on $V_G$.  This was the
motivation of Yacoby {\it et al.} \cite{yacoby}, who used a
\textbf{closed mesoscopic ABI}, where a ring made of the two paths
was connected to two terminals leading to two electron reservoirs.
Indeed, the measured conductance was periodic in $\phi$, and the
detailed dependence of $\textbf T$ on $\phi$ varied with $V_G$,
showing resonances. However, close to a resonance the data did not
fit the simple 2-slit formula; they required more harmonics in
$\phi$. The data also exhibited ``phase rigidity": the fitted
phase $\beta$ did not follow the continuous variation with $V_G$
(as would be implied from the 2-slit scenario, where
$\beta=\alpha_{QD}+{\rm const}$). Instead, $\beta$ exhibited
discrete jumps by $\pm \pi$ as $V_G$ passes through each
resonance.

This ``phase rigidity" results from the Onsager relations.  Unlike
the 2-slit geometry, the closed ABI requires many reflections of
the electron waves from the `forks' connecting the ring with the
leads. Each such reflection adds a term to the interference sum of
amplitudes, and modifies the simple 2-slit formula. In fact,
unitarity (conservation of current) and time reversal symmetry
imply that the conductance ${\textbf G}$ (and therefore also the
transmission ${\textbf T}$) obey the symmetry ${\textbf
G}(\phi)={\textbf G}(-\phi)$ \cite{onsager,rigid}, and therefore
$\beta$ {\it must} be equal to $0$ or $\pi$. As discussed below,
the additional reflections also explain the need for higher
harmonics near resonances.

 Later experiments \cite{schuster} used an \textbf{open interferometer},
by adding `lossy' channels which break unitarity. Indeed, fitting
the conductance to Eq. (\ref{twoslit}) yielded a phase
$\beta(V_G)$ which was interpreted as representing the `intrinsic'
$\alpha_{QD}(V_G)$. Below we discuss the applicability of the
two-slit formula to such experiments \cite{bih}.

In all these experiments, one is restricted to small mesoscopic
systems and to low temperatures, in order to maintain the
coherence of the wave functions and observe quantum interference
\cite{book}. The above experiments were performed at relatively
high temperatures, where the main effect of the Coulomb
interactions arises via Coulomb blockade, which introduces a
separate resonance whenever $V_G$ allows the addition of one more
electron to the QD. As the temperature $T$ decreases below the
Kondo temperature $T_K$, and the QD is occupied by an odd number
of electrons, the spin of these electrons is dynamically screened
by the electrons in the Fermi sea, yielding a large conductance
 ${\textbf G}$ through the QD, close to the unitarity value $2e^2/h$,
and a transmission phase $\alpha_{QD}$ equal to $\pi/2$
\cite{hewson}. Aiming to test these predictions then led to
experiments using the ABI at very low $T$, for both a closed
(``two-terminal") ABI \cite{vander} and an open
(``multi-terminal") ABI \cite{ji}. Both experiments exhibited the
Aharonov-Bohm oscillations with $\phi$. The former experiments
exhibited the expected ``phase rigidity", but there has been no
quantitative analysis of these data. The latter experiments used
the two-slit formula to ``measure" the transmission phase $\beta$,
and found an unexpected variety of behaviors which were
inconsistent with the theoretical value $\alpha_{QD}=\pi/2$.

During the last few years we obtained several theoretical results
concerning the interpretation of the above experiments. Below we
review some of these earlier results, and report on some recent
new results.

\section{Model for the ABI}\label{sec2}

All our calculations are done for the model shown in Fig. 1. The
conductance ${\textbf G}$ is measured between the two leads which
are attached to sites ``L" and ``R" on the ABI ring. The QD
(denoted ``D") is on the upper branch, and is connected to L (R)
via $n_l$ ($n_r$) sites. The lower ``reference" branch contains
$n_0$ sites. Except for the QD, we use a tight binding model, with
the real hopping matrix elements as indicated in the figure, and
with site energies $\epsilon_l,~\epsilon_r$ and $\epsilon_0$ on
the respective branches, and  $\epsilon_L,~\epsilon_R$ on sites L
and R. The site energies on the leads are set at 0. The normalized
flux $\phi$ is introduced as a phase factor in
$J_{D1}=J_{1D}^\ast=j_l e^{i\phi}$ (gauge invariance allows
placing it on any bond(s) around the ring). Assuming that the
transmission is dominated by a single resonance on the QD, the
Hamiltonian on the dot is assumed to have the form
\begin{eqnarray}
{\textbf H}_d=\epsilon_d\sum_\sigma
n_{d\sigma}+Un_{d\uparrow}n_{d\downarrow},
\end{eqnarray}
with $n_{d\sigma}=d^\dagger_\sigma d_\sigma$ being the number
operator for an electron with spin $\sigma$ on the QD.  We also
assume that $U$ is very large, and ignore the resonance at
$2\epsilon_d+U$. For the open ABI, each dashed line represents an
additional lead, with a hopping matrix element $-J_X$ on its first
bond \cite{bih}.

\begin{figure}[tb]
  \begin{center}
    \vspace{-.2cm}\includegraphics[angle=0,width=.5\textwidth]{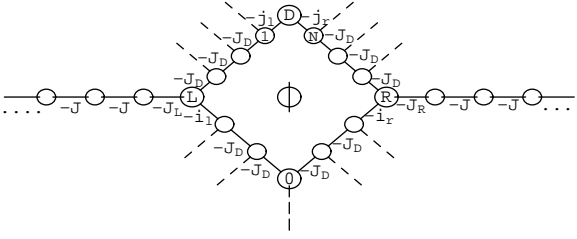}
    \caption{Model for the ABI.}
    \label{fig:randWalk}
  \end{center}
\end{figure}

In the absence of the lower ``reference" branch, and when
$J_L=J_D=J_R=J$, $\epsilon_L=\epsilon_R=\epsilon_l=\epsilon_r=0$,
the transmission amplitude through the upper branch reduces to
\cite{ng}
\begin{eqnarray}
t_{QD}&=&-i \gamma_D\sin\alpha_{QD}e^{i\alpha_{QD}}\nonumber\\
&=&2i\sin|ka|j_lj_rG_{up}(\epsilon_k)/J,\label{tQD}
\end{eqnarray}
where $\epsilon_k=-2J\cos(ka)$ is the energy of the electron in
the band (equal to the Fermi energy), and $G_{up}$ is the retarded
Green function of an electron on the QD for this geometry. Here,
$\gamma_D=2j_lj_r/(j_l^2+j_r^2)$. If these one-dimensional
conditions apply, then this equation dictates a relation between
the magnitude and the phase of $t_{QD}$:
\begin{eqnarray}
{\textbf T}_{QD}=\gamma_D^2\sin^2\alpha_{QD}.\label{sin2}
\end{eqnarray}

Without interactions,
\begin{eqnarray}
G_{up}=1/[\epsilon_k-\epsilon_d-\Sigma_0],\label{Gnon}
\end{eqnarray}
with the self-energy $\Sigma_0(\epsilon_k)$ depending on details
of the leads connected to the QD.

\section{The closed ABI}\label{sec3}

In Ref. \cite{prl2}, we used the equation-of-motion method to
calculate the transmission amplitude through the closed ABI, $t$,
for a simple version of Fig. 1: $n_l=n_r=0,~n_0=1$. We ended up
with the result
\begin{eqnarray}
t=A_Dt_{QD} e^{i \phi}+A_B t_B, \label{tt1}
\end{eqnarray}
where $t_B$ was the ``bare" transmission amplitude through the
lower ``reference" path, when the upper path was disconnected, and
the coefficients $A_D$ and $A_B$ contain all the additional
processes in which the electron ``visits" the reference site
(denoted 0 in the figure), or the dot, respectively. Without
interactions, we were able to prove that the $\phi$-dependence of
${\textbf T}$ has the form
\begin{eqnarray}
{\textbf T}={\bf C}\frac{1+K^2+2
K\cos\phi}{1+2P(z+\cos\phi)+Q(z+\cos\phi)^2}, \label{ttt}
\end{eqnarray}
where the five coefficients ${\bf C},~K,~P,~Q$ and $z$ are all
real, and independent of $\phi$.  These parameters do depend
explicitly on both the real and imaginary parts of $G_{QD}$.
Although the numerator in Eq. (\ref{ttt}) looks like the 2-slit
formula, with $\beta=0$ or $\pi$ (depending on ${\rm sign} K$),
 the new physics, related to the many ``trips" of the electron
 around the ring, is contained in the denominator. Note that Eq.
 (\ref{ttt}) obeys the
``phase rigidity", and also requires many harmonics in the flux.
 A 5-parameter fit to the explicit
$\phi$-dependence in Eq. (\ref{ttt}) for given values of $V_G$ and
of the other ABI parameters then allows one to extract the
$V_G$-dependence of $G_{QD}$.

Reference \cite{prl2} also contained some speculations on how to
use Eq. (\ref{ttt}) in the presence of interactions, when the
corrections to the self-energy due to the lower branch can be
expanded in a Fourier series in $\phi$, and when this series is
dominated by the first harmonic (thus renormalizing the parameter
$z$ above). More recently we  generalized Eq. (\ref{ttt}) to the
strongly correlated Kondo case \cite{intABI}. Deep inside the
Kondo regime, when the QD is occupied by a single electron, the
Fermi liquid conditions constrain the Green function on the QD.
The resulting Green function then depends only on  the {\it
non-interacting} self-energy $\Sigma_0(\epsilon_k)$ [see Eq.
(\ref{Gnon})]. This allowed us to obtain the exact dependence of
the conductance through the ABI on the flux $\phi$,
\begin{eqnarray}
{\textbf
G}=\frac{A+B\cos\phi+C\cos^2\phi}{1+D\cos\phi+E\cos^2\phi}.\label{kondo}
\end{eqnarray}
In this limit ($T \rightarrow 0,~\epsilon_d \rightarrow -\infty,~U
\rightarrow \infty$), all the coefficients depend only on the
non-interacting parts of the ABI. This conductance reaches the
unitarity limit $2e^2/h$ for some fluxes $\phi$, but its
dependence on $\phi$ contains a non-trivial structure which is
specific to the particular geometry, and not to the physics of the
QD itself. Using a recently derived approximate solution for the
Green function in the Kondo regime, for $U\rightarrow \infty$
\cite{kondo}, we found that Eq. (\ref{kondo}) presents an
excellent fit to the ``data" for all $\epsilon_d$ and $T \ll T_K$.
Fits to such data again allow the extraction of the Green function
on the dot from the closed ABI data, also in the Kondo regime.

\section{The open ABI}

Imitating the experiments of Schuster {\it et al.}
\cite{schuster}, we introduce the opening of the ABI via the side
branches which are attached to all the sites on the ring except
the QD. The algebra remains similar to the case of the closed ABI,
except that various sites now have a complex self-energy, due to
the side branches. Since the data were taken in the Coulomb
blockade, each resonance can be imitated by a separate level on
the QD, with the Coulomb interaction representing the distance
between such levels. For the non-interacting case, we again end up
with an equation like (\ref{ttt}), except that the numerator is
now replaced by $|1+Ke^{i\phi}|^2$, and $K$ is now {\it a complex
number}. Thus, the numerator assumes the form $1+|K|^2+2\Re[K
e^{i\phi}]$, similar to the 2-slit Eq. (\ref{twoslit}). However,
the coefficients, as well as the imaginary part of $K$, depend on
the parameter $J_X$ which is a measure for the opening. In Ref.
\cite{bih} we fitted these calculated results  to the 2-slit
formula (\ref{twoslit}), and discovered that the ``measured" phase
$\beta$ reproduces the correct phase $\alpha_{QD}$ only when the
structure of the ABI and of the side branches are optimized: to
have $\beta=\alpha_{QD}$, the electron must cross each branch only
once, with no reflections anywhere. One necessary condition for
this was appreciated qualitatively in an earlier publication
\cite{prl1}:  the electron must practically never be reflected
from the ``forks" where the ring meets the incoming and outgoing
leads. A second condition requires that the electron also passes
through the QD only once, and does not ``reverberate" back and
forth between the ``combs" on its two sides. In our model, both
conditions are achieved by having a very small net transmission
through {\it and} a very small reflection from each ``comb" of
``lossy" channels. As $J_X$ increases, the transmission through
the ``lossy" scatterers decreases, but the reflection from them
increases. Therefore, there is only an {\it intermediate} range of
$J_X$ where $\beta=\alpha_{QD}$.

In principle, experimentalists should vary the strength of the
coupling to the side ``lossy" branches, is search for the optimal
intermediate range. However, at the moment we only have the data
published in Ref. \cite{schuster}, taken from a single realization
of the open ABI.  We now present two criteria which can check the
consistency of the two-slit conditions in these experiments. One
such criterion follows from Eq. (\ref{selfcon}), which requires
that
\begin{eqnarray}
b\equiv\Bigl (\frac{B}{\max[B]}\Bigr )^2 = |t_1|^2=
\frac{A-\min[A]}{\max[A]-\min[A]}\equiv a.\label{test1}
\end{eqnarray}
A second criterion follows from Eq. (\ref{sin2}):
\begin{eqnarray}
a=b=|t_1|^2=\gamma_D^2
\sin(\alpha_{QD})=\gamma_D^2\sin^2\beta.\label{test2}
\end{eqnarray}
Taking the data for $A,~B$ and $\beta$ from the graphs of Ref.
\cite{schuster}, we check these criteria in Fig. 2. We adjusted
the scale of $b$ arbitrarily, to optimize the fit between the
curves. The reader can now judge if the data obey these criteria.
At least over some energy range, and remembering the
uncertainties, the data may be reasonably consistent with the
two-slit requirements. However,  a systematic study of open ABI's
is still desired.

\begin{figure}[tb]
  \begin{center}
    \includegraphics[angle=0,width=.5\textwidth]{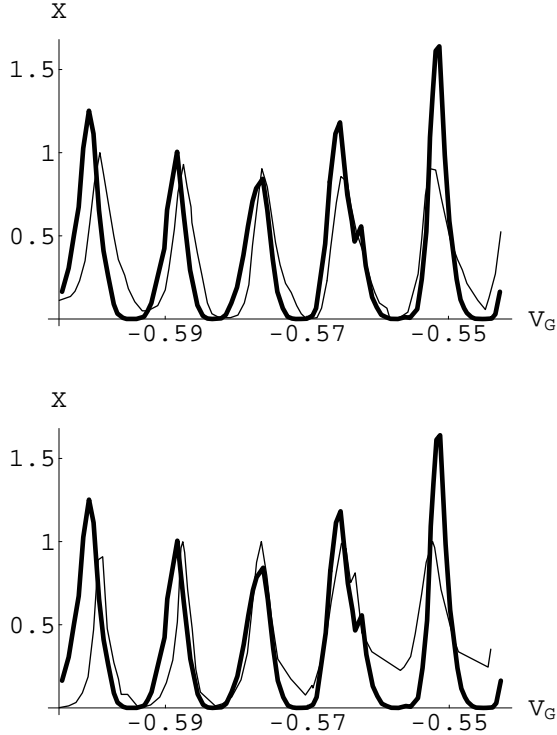}
    \caption{Tests of Eqs. (\ref{test1})
    (\ref{test2}). The thick lines represent $b$
    (rescaled arbitrarity), while
    the thin lines represent $a$ (top panel) and
    $\sin^2[\pi(\beta-.03)]$ (the data in Ref. \cite{schuster}
    give $\beta$ in units of $\pi$).}
    \label{tests}
  \end{center}
\end{figure}

More recently, we have also analyzed the open ABI for strongly
correlated dots \cite{intABI}. Similar to the closed case, deep
inside the Kondo regime the numerator in Eq. (\ref{kondo}) is now
replaced by $A+B\cos(\phi+\beta)+C\cos(2\phi+\gamma)$, with the
parameters depending only on the non-interacting parts. Thus,
$\beta$ depends strongly on these parts of the ABI, and
experiments may end up with arbitrary values of this ``measured"
phase, instead of the Kondo value of $\alpha_{QD}=\pi/2$. This may
explain the unexplained observations in Ref. \cite{ji}.
Furthermore, we find that $T_K$ depends strongly on the flux and
on $J_X$, and therefore the opening of the ABI may in fact destroy
the Kondo correlations on the QD and eliminate the Kondo regime
altogether. Thus, the measurements in the Kondo regime are much
more sensitive to the details of the ABI, compared to the
non-interacting case.

\section{Concluding remarks}

Our main conclusion is that opening the ABI introduces many
undesired uncertainties into the analysis of the data. It also
reduces the outcoming current, due to the losses on the side
branches. In contrast, we now have a reasonable quantitative
understanding of the conductance through the closed ABI. It would
therefore be useful to  have new systematic experimental studies
of closed ABI's, which could test our various quantitative
predictions.

\vspace{.5cm}

\noindent{\bf Acknowledgements}

\vspace{.2cm}
 We thank B. I. Halperin, Y. Meir and
 A. Schiller for helpful conversations. This project was carried out in
a center of excellence supported by the Israel Science Foundation
(under grant \#1566/04), with additional support from the Albert
Einstein Minerva Center for Theoretical Physics at the Weizmann
Institute of Science, and from the German Federal Ministry of
Education and Research (BMBF) within the Framework of the
German-Israeli Project Cooperation (DIP). Work at Argonne
supported by the U. S. department of energy, Basic Energy
Sciences--Materials Sciences, under contract \#W-31-109-ENG-38.

\end{document}